\documentclass{sig-alternate-05-2015}
\usepackage{epsfig}
\usepackage{graphicx,dblfloatfix}
\usepackage{tabularx}
\usepackage{amsmath}
\usepackage{amssymb}
\usepackage{relsize} 
\usepackage{wasysym}
\usepackage{graphicx}
\usepackage{tikz}
\usetikzlibrary{calc,positioning,shapes.geometric}
\usetikzlibrary{trees,shapes}
\usepackage{subfig}
\usepackage{float}
\usepackage{algorithm}
\usepackage{algpseudocode}
\usepackage{pifont}
\usepackage{color}
\usepackage{colortbl}
\usepackage{multirow}
\usepackage{url}

\begin{document}
\title{A Comparative Study of Different Source Code Metrics and Machine Learning Algorithms for Predicting Change Proneness of Object Oriented Systems}
\numberofauthors{2}
\author{
	\alignauthor {Lov Kumar}\\ 
	\affaddr{BITS Pilani Hyderabad, India}  \\
	\email{lovkumar505@gmail.com} \\
	\alignauthor Ashish Sureka\\   
	\affaddr{Ashoka University, India} \\
	\email{ashish.sureka@ashoka.edu.in} 
}
	
	\CopyrightYear{2018} 
	\setcopyright{acmcopyright}
	\conferenceinfo{ISEC '18,}{February, 2018,  Hyderabad, India}
	\isbn{111-1-1111-1111-1/11/11}\acmPrice{\$15.00}
	\doi{http://dx.doi.org/11.1111/1111111.11111111}
	
	\maketitle
\begin{abstract}
Change-prone classes or modules are defined as software components in the source code which are likely to change in the future. Change-proneness prediction is useful to the maintenance team as they can optimize and focus their testing resources on the modules which have a higher likelihood of change. Change-proneness prediction model can be built by using source code metrics as predictors or features within a machine learning classification framework. In this paper, twenty one source code metrics are computed to develop a statistical model for predicting change-proneness modules. Since the performance of the change-proneness model depends on the source code metrics, they are used as independent variables or predictors for the change-proneness model. Eleven different feature selection techniques (including the usage of all the $21$ proposed source code metrics described in the paper) are used to remove irrelevant features and select the best set of features. The effectiveness of the set of source code metrics are evaluated using eighteen different classiffication techniques and three ensemble techniques. Experimental results demonstrate that the model based on selected set of source code metrics after applying feature selection techniques achieves better results as compared to the model using all source code metrics as predictors. Our experimental results reveal that the predictive model developed using LSSVM-RBF yields better result as compared to other classification techniques
\end{abstract}


\section{Introduction}
Change-prone classes or modules are defined as software components in the source code which are likely to change in the future. Prediction and early identification of such components are useful to the maintenance team as they can optimize and focus their testing resources on the modules which have a higher likelihood of change. Prediction of change prone components is an area which has attracted several researchers attention \cite{Kumar2017SIGSOFT}\cite{Kumar2017ISEC}\cite{Kumar2017ISECIND}\cite{kumar2017MALT}\cite{sharafat2008change}\cite{lu2012ability}. Building effective and accurate change-proneness predictive models is a technically challenging problem. Change-proneness prediction models are generally developed using structural measurement of software (software metrics) i.e, size, cohesion, coupling, and inheritance \cite{chen2009empirical}\cite{chidamber}\cite{henry}\cite{Basili}\cite{malhotra2012fault}. Source code metrics are used to measure the internal structure of software system such as complexity, coupling, cohesion, inheritance, and size.  In this work, $21$ software metrics are considered to develop a model for predicting change-proneness modules. 

Since our change-proneness prediction model is based on source code metrics, the selection of the suitable set of source code metrics becomes an integral component of the model development process. Selection of right set of features or metrics is an important data pre-processing task while building machine learning based classifiers. In the work presented in this paper,  eleven different feature selection techniques are used to validate the source code metrics and identify suitable set of source code metrics with the aim to reduce irrelevant or non-informative metrics and thereby improve the performance of change-proneness prediction model. The effectiveness of these set of source code metrics are evaluated using eighteen different learning algorithms and three ensemble techniques \cite{Kumar2017SIGSOFT}\cite{Kumar2017ISEC}\cite{Kumar2017ISECIND}\cite{kumar2017MALT}\cite{kumar2017compsac}.\\ 

\noindent \textbf{Research Aim:} The research aim of the work presented in this paper is to investigate the application of twenty one source code metrics, eleven feature different feature extraction or selection methods and three different ensemble methods for predicting change proneness. The number and type of source code metrics, feature selection methods and machine learning techniques in this study is unexplored and forms the novel and unique research contributions of the work. 

\section{Experimental Dataset}	
We conduct experiments on open-source publicly available dataset so that our experiments can be easily reproduced and the results can be replicated. Table \ref{cP} displays a list of 10 Eclipse $2.0$ and Eclipse $2.1$ plug-ins used in our experiments. The source code for the Eclipse source build 2.0 is downloaded from the URL\footnote{\url{https://goo.gl/bFM30Y}} as a zip file and similarly the source code for the the Eclipse source build $2.0$ is downloaded from the URL\footnote{\url{https://goo.gl/3z33Rw
}} as a zip file. All the $10$ plug-ins are Java based applications and are present in both the versions of the Eclipse. We compute the object oriented metrics, the number of classes and the number of source lines of code changes for each Java class in Eclipse $2.0$ version which also appears in Eclipse $2.1$ version. 

The source code metrics are computed using a tool called as Understand from Scitools\footnote{\url{https://scitools.com/}}. We compute $21$ metrics listed on the website\footnote{\url{https://scitools.com/support/metrics_list/}} of the Understand tool. The metrics cover a diverse range of source code properties such as lines of code, cyclomatic complexity, coupling between objects, class methods, class variables, functions, instance methods and variables and depth of inheritance tree. We use a diff tool called as Jar Compare\footnote{\url{http://www.extradata.com/products/jarc/}} for computing differences between Class files in Java JAR archives. The tool is used for comparing changes between software builds or releases and we use it to identify the Classes which were changed between the two versions within a plug-in in our experimental dataset. Table \ref{cP} reveals that the largest plug-in in our dataset is JDT having $1943$ classes in Eclipse version $2.0$ and number of changed classes as $1221$. The number of changed classes vary from a minimum of $31$ to a maximum of $1221$. The percentage of changed classes vary from $29.81\%$ to a maximum of $67.67\%$.  
\begin{table}[h]
	\centering
	\caption{Experimental Dataset Details (DS: Dataset, CHN: Changed)}
	\label{cP}
	
	\begin{tabular}{|l|l|c|c|c|c|}
		\hline
		\textbf{ID} & \textbf{Name} & \textbf{\# class} & \textbf{\# CHN} &  \textbf{\% CHN} \\ \hline
		DS1 & compare & 83 & 38 & 45.78 \\ \hline
		DS2 & webdav & 104 & 31 & 29.81 \\ \hline
		DS3 & debug & 133 & 90 & 67.67 \\ \hline
		DS4 & update & 249 & 167 & 67.07 \\ \hline
		DS5 & core & 250 & 90 & 36.00 \\ \hline
		DS6 & swt & 344 & 126 & 36.63 \\ \hline
		DS7 & team & 372 & 236 & 63.44 \\ \hline
		DS8 & pde & 487 & 269 & 55.24 \\ \hline
		DS9 & ui & 826 & 516 & 62.47 \\ \hline
		DS10 & jdt & 1943 & 1221 & 62.84 \\ \hline
	\end{tabular}
\end{table}
\section{Research Questions (RQ)}
We frame several research questions and conduct empirical analysis to answer the stated research questions. Following are the list of research questions. \\

\noindent \textbf{RQ1:} \textit{Are the $21$ source code metrics able to predict change-proneness of object-oriented software ?}\\

\noindent Our objective is to investigate the relationship between each source code metrics and change-proneness. We apply the Wilcoxon signed rank test and Univariate Logistic Regression (ULR) to determine the correlation between a particular metric and change-proneness of classes. \\

\noindent \textbf{RQ2:} \textit{Does the selected set of source code metrics are better to predict whether a class is change-proneness or not ?} \\
	
\noindent This step aims to evaluate the performance of selected set of metrics. In this study, four steps i.e., Wilcoxon signed rank test, ULR, Correlation analysis, and forward stepwise selection procedure are followed  have been considered for finding subset of source code metrics which are better to predict whether there is change-proneness or not. \\
	
\noindent \textbf{RQ3:} \textit{What is the variation in performance (measured in-terms of accuracy and F-measure) of different classifiers models over different set of source code metrics?} \\
	
\noindent This question helps to investigate the performance of different classifiers for change-proneness techniques. \\
	
\noindent \textbf{RQ4:} \textit{Which feature selection method works best for predicting change-proneness of object-oriented software?} \\
	
\noindent The performance of feature selection method is based on the nature of the change-proneness dataset. Here two different performance parameters have been considered to compare different feature selection methods. \\
	
\noindent \textbf{RQ5:} \textit{Does the ensemble methods improve the performance of the change-proneness prediction models?} \\
	
\noindent This question helps to investigate the performance of different types of ensemble methods.  In the present work, we have considered heterogeneous ensemble method with three different combination rules (2 Linear, and 1 nonlinear). \\
	
\noindent \textbf{RQ6:} \textit{Does the feature selection techniques effect the performance of the classification techniques ?} \\
	
\noindent This question investigates the variation of performance of classification method over different feature selection techniques. It may possible that some feature selection techniques works very well with specific classification method. Thus,  in this study, different feature selection techniques are evaluated using twenty one different classification methods. 


\section{Experimental Setup and Results}
\label{resultoos}
Our solution approach consists of multiple steps. We begin by creating the dataset containing source code metrics and change-proneness classes of object-oriented software. Then, we apply dimensionality reduction and remove irrelevant features. In our source code metrics validation framework  we compare all metrics, five feature ranking techniques, and five feature subset selection techniques. These $12$ different set of source code metrics are considered as input to develop a model using eighteen different learning algorithms and three ensemble techniques. In this work, 12 sets of source code metrics, 21 classifiers have been considered to developed change-proneness models and evaluate the performance of all the combinations resulting in a comprehensive and in-depth experimental evaluation. 

In our experiments, the standard technique of $5$ fold cross-validation has been considered for the purpose of evaluating and then comparing the predictive models. Cross-validation approach is employed to assess and compare the statistical models by partitioning or segmenting the dataset into two portions called as training and test datasets \cite{kohavi}.  The training dataset segment of the divided subset is used to learn the model and the remaining data is used to validate the model accuracy. In K-fold cross-validation technique, the model building dataset is first partitioned into $K$ equal (or roughly equal) sized partitions called as the folds \cite{kohavi}. $K$-1 folds are used for training purpose and the rest $1$ fold is used for testing for the final goal of creating each of the $K$ models. The advantage of K-fold-cross-validation lies in its ability to utilize a single dataset for both training and testing and averaging the results across multiple partitions by removing bias. In our study, $5$-fold cross-validation has been applied for model building and comparison. The performance of developed model is evaluated using two different performance parameters such as Accuracy (\%) and F-Measure. Finally, statistical tests have been conducted  to identify the best performing change-proneness prediction model.

In our experiments, eleven different types of feature selection methods are applied to eliminate some of the irrelevant and redundant original variables to increase the training speed and accuracy of the classifier. These selected set of source code metrics are used as input of the change-proneness prediction models.  The eleven different types of feature selection techniques that are used in this study are 
\textbf{(1)} Proposed source code metrics validation framework (PFST), five feature ranking techniques such as \textbf{(2)} Chi Squared test (FR1), \textbf{(3)} Gain Ratio Feature Evaluation (FR2), \textbf{(4)} oneR Feature Evaluation (FR3), \textbf{(5)} info gain feature evaluation (FR4), \textbf{(6)} principal component analysis (PCA) (FR5), and five feature subset selection techniques \textbf{(7)} correlation based feature selection (CFS) technique (FS1), \textbf{(8)} consistency feature selection (FS2), \textbf{(9)} filtered subset evaluation (FS3), \textbf{(10)} rough set analysis (RSA) (FS4), and \textbf{(11)} genetic algorithm (FS5). These feature selection techniques provides us guidance on selection a subset of the original features which are useful in developing a good estimator or predictor of change-proneness.

\subsection{Source Code Metrics Validation Framework}
The proposed source code metrics validation framework has been applied to suitable set of source code metrics for change-proneness prediction. In this Section, we present the detailed description of the selection of source code metrics at each step of proposed framework. Initially, 
filter approach is applied on $21$ source code metrics to remove insignificant features. After computation of significant set of source code metrics, wrapper approach has been applied to select best set of metrics for object-oriented change-proneness prediction.  The detailed description of each steps are summarized in subsequent sections.

\subsubsection{Wilcoxon signed rank test}
\begin{figure*}[h]
	\centering
	\includegraphics[width=14.5cm, height=4.20cm]{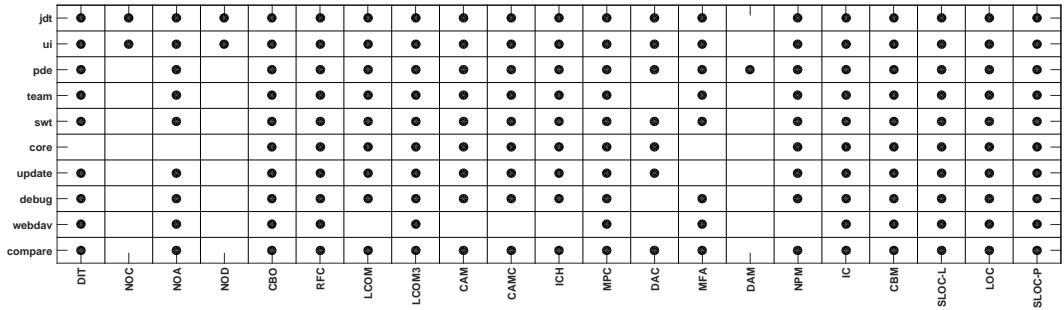}
	
	\caption{Wilcoxon signed rank test results}
	\label{t-testoos}
\end{figure*}

Initially,  Wilcoxon signed rank test is applied to evaluate the relationship and individual effect of each of the $21$ source code metrics on the change-proneness of classes. The objective of this step is to identify metrics which are significantly, moderately and not related to the change-proneness of classes. The $21$ source code metrics represents independent variables and the dependent variable represents the change value for each class (can take only of the two values: changed or not changed). Here, Wilcoxon signed rank test is applied on each source code metric and their $p-value$ is considered to measure \textit{how effectively it separates the change or non-change-proneness groups}. The result of Wilcoxon signed rank test analysis is shown in Figure \ref{t-testoos}.

The $ p-value $ are represented using two different symbols (blank box ($\Box$): $ p-value >$ 0.05 and box with black circle ($\CIRCLE$): $p-value <=$ 0.05). The metrics having $p-value$ lesser than $0.05$ are significant differentiators of the change or non-change-proneness classes. From Figure \ref{t-testoos}, it can infer that DIT, NOA, CBO, RFC, LCOM, LCOM3, CAM, CAMC, ICH, MPC, DAC, MFA, NPM, IC, CBM, CLOC-L, LOC, AND SLOC-P source code metrics significantly differentiate the change or non-change-proneness classes for compare data. Therefore, we conclude that these metrics are significantly differentiators of the change or non-change-proneness classes.

In this work, error box plot diagrams have been considered to cross-check the results of Wilcoxon signed rank test analysis. 
A source code metric statistically differentiate between the change-proneness and non-change-proneness groups if and only of their mean of 95\% confidence intervals do not overlap otherwise this metric is not significant metrics for change-proneness prediction. We observe that the mean of 95\% confidence intervals of change and non change-proneness classes using DIT, NOA, CBO, RFC, LCOM, LCOM3, CAM, CAMC, ICH, MPC, DAC, MFA, NPM, IC, CBM, CLOC-L, LOC, and SLOC-P metrics do not overlap (i.e. are significantly different). We conclude that these metrics are capable of identifying the change or non-change proneness classes.

\subsubsection{Univariate Logistic Regression (ULR) Analysis:}
\begin{figure*}[h!]	
	\centering
	\includegraphics[width=14.5cm, height=4.0cm]{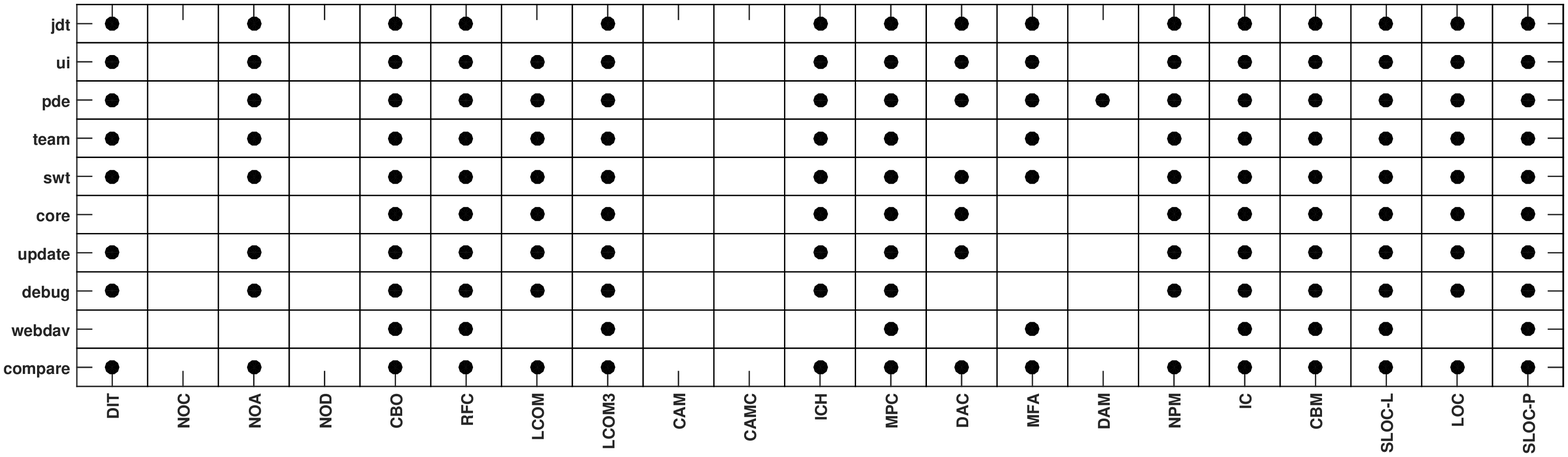}
	\caption{Univariate Logistic Regression (ULR) Analysis:}
	\label{ulroos}
\end{figure*}

Univariate logistic regression (ULR) analysis is applied on selected set of source code metrics using Wilcoxon signed rank test to investigate whether the selected set of metrics using $ Wilcoxon signed rank test $ analysis are significant predictors of change-proneness classes or not. Univariate logistic regression helps in computing the percent or extent of variance (a predictor of statistical relationship between two variables) in the dependent variable explained by the independent variables. The selected set of source code metrics represents independent variables and the dependent variable represents the change value for each class (can take only of the two values: changed or not changed). A source code metrics is significant predictor of class change-proneness if its p-value of coefficient is less than $0.05$. The source code metrics having $p-value$ values of coefficient lesser than $0.05$ are shown using box with black circle ($\CIRCLE$) in Figure \ref{ulroos}.

\begin{figure*}[t]	
	\centering
	\includegraphics[width=15.5cm, height=4.80cm]{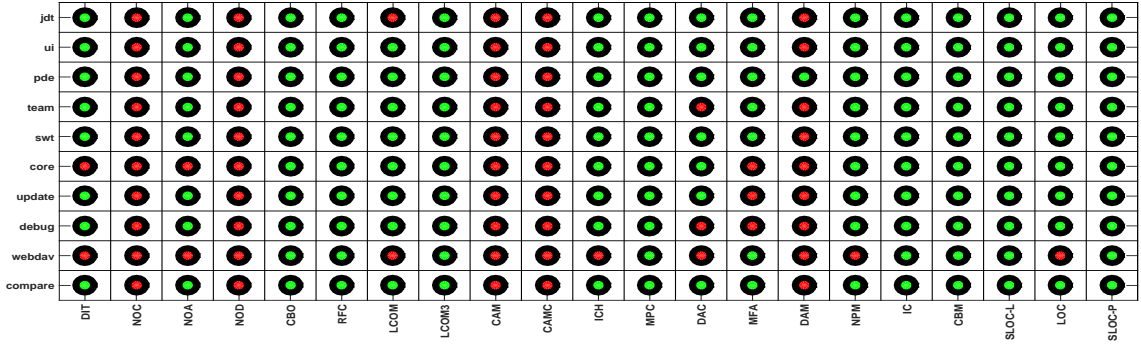}
	\caption{Hypothesis}
	\label{hypothesisoos}
\end{figure*}

Figure \ref{ulroos} reveals that the coefficient p-value of metrics NOC, NOD, CAM, CAMC, and DAM is greater than the commonly used alpha threshold or level of $0.05$ and hence they are not statistically significant predictors. From ULR analysis, it has been 
infer that we should remove metrics NOC, NOD, CAM, CAMC, and DAM  from model building. From Figure \ref{ulroos}, it has been also observed that these metrics DIT, NOA, CBO, RFC, LCOM, LCOM3, ICH, MPC, DAC, MFA, NPM, IC, CBM, CLOC-L, LOC, and SLOC-P have a low p-value value ranging between $0$ and $0.05$ and hence they are useful predictors for change-prone estimator The acceptance and rejection of hypotheses for all considered datasets are shown in  
Figure \ref{hypothesisoos} using green circle (\textcolor{green}{$\CIRCLE$}) and red circle (\textcolor{red}{$\CIRCLE$}) respectively. The Figure \ref{hypothesisoos} reveals that, $16$ source code metrics are the good predictor for change-proneness prediction of compare dataset.

\subsubsection{Cross Correlation Analysis:}
\begin{figure}[H]		
	\centering
	\includegraphics[width=8cm, height=6.0cm]{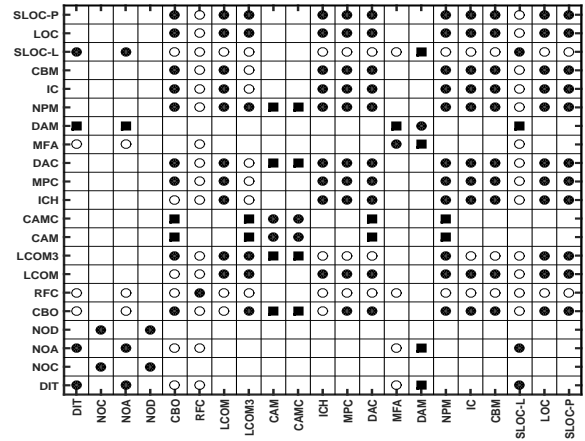}
	\caption{Correlation between source code metrics}
	\label{corroos}
\end{figure}

In this work, the association between different pairs of source code metrics are computed using Pearson's correlations coefficient ($r$).
The coefficient of correlation $r$ measures the strength and direction of the linear relationship between two variables. Figure \ref{corroos} displays the correlation results for compare dataset on correlation analysis between the $21$ metrics. The results of other datasets are of similar types.  In Figure \ref{corroos}, a Black circle represents an $r$ value between $0.7$ and $1.0$ or between $-0.7$ and $-1.0$ indicating a strong positive or negative linear relationship respectively. A white circle $r$ value between $0.3$ and $0.7$ or $-0.3$ and $-0.7$ indicating a weak positive or negative linear relationship respectively. A blank cell represents no linear relationships between the two variables. For example, based on Figure \ref{corroos}, it can be seen that there is a strong positive linear relationship between LCOM and seven other variables LCOM3, ICH, MPC, DAC, NPM, IC, CBM, LOC and SLOC-P. On the other hand, it is also observe a weak linear relationship between LCOM and COB, RFC, and SLOC-L.  Figure \ref{corroos} reveals association between different suite of metrics and not just associations between metrics within the same suite. 

\begin{figure*}[h!]
	\centering
	\includegraphics[width=14.5cm, height=4.5cm]{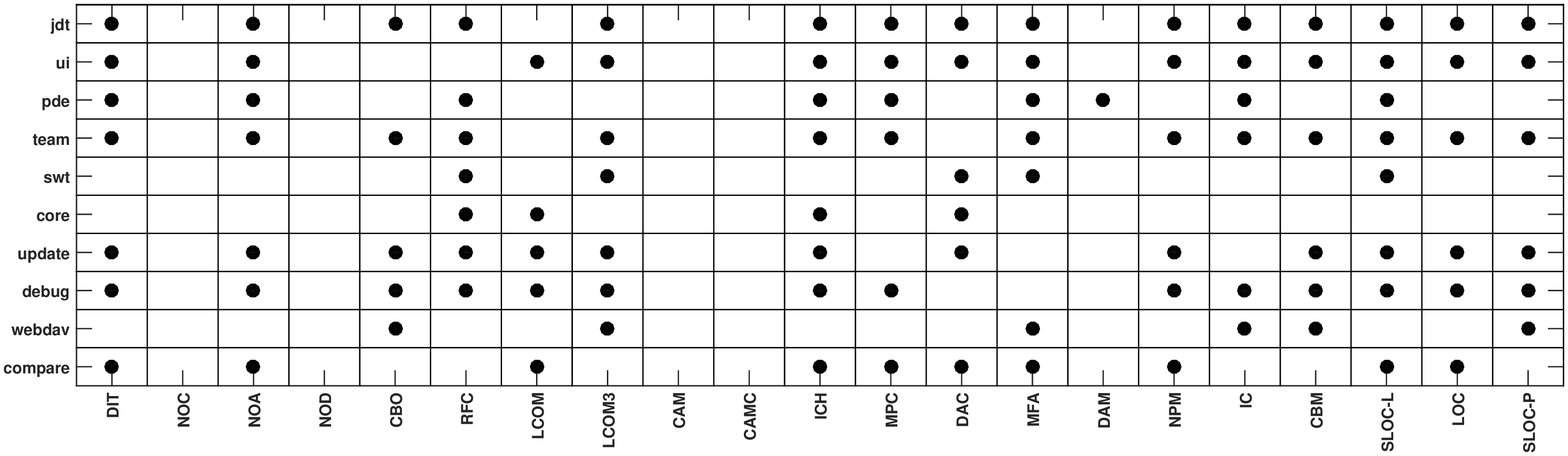}
	\caption{Cross Correlation Analysis}
	\label{ccorroos}
\end{figure*}

In this work, cross correlation analysis are performed on  significant source code metrics identified using Wilcoxon signed rank test and ULR analysis. If a significant source code metric shows higher correlation i.e., r-value $>=$0.7 or r-value $<=$-0.7  with other significant source code metrics then we check the performance of these source code metric individually and in the combine basis for change-proneness prediction and select a metric or group of metrics, whomsoever perform better. The selected set of source code metrics after cross correlation analysis are shown in Figure \ref{ccorroos}. The selection of source code metrics are depicted using black circle ($\CIRCLE$). 

\subsubsection{Multivariate Linear Regression Stepwise Forward Selection:}

\begin{figure*}[h!]
	\centering
	\includegraphics[width=14.3cm, height=4.0cm]{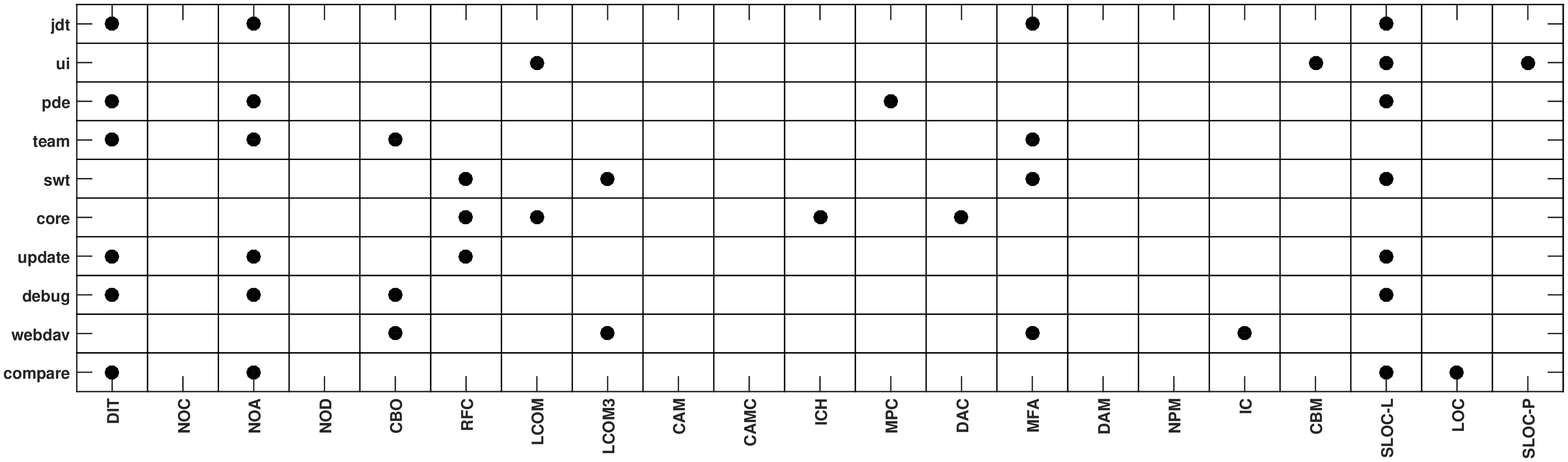}
	\caption{Multivariate Linear Regression Stepwise Forward Selection}
	\label{step4oos}
\end{figure*}

Eliminating the insignificant metrics and collinearity
among the metrics does not necessarily mean that
we have the right set of source code metrics for change-proneness
prediction. To select right set of source code
metrics, multivariate linear regression stepwise
forward selection process has been considered. The best set of source code metrics after all four analysis i.e., Wilcoxon signed rank test, ULR analysis, Cross Correlation analysis and multivariate linear regression stepwise forward selection method are shown using black circle ($\CIRCLE$) in Figure \ref{step4oos}. Form Figure \ref{step4oos}, it is observed that the set of metrics selected using proposed source code metrics validation framework are DIT, NOA, SLOC-L, and LOC for compare dataset.

\subsubsection{Over All}
\begin{figure*}[h!]
	\centering
	\includegraphics[width=15.5cm, height=6cm]{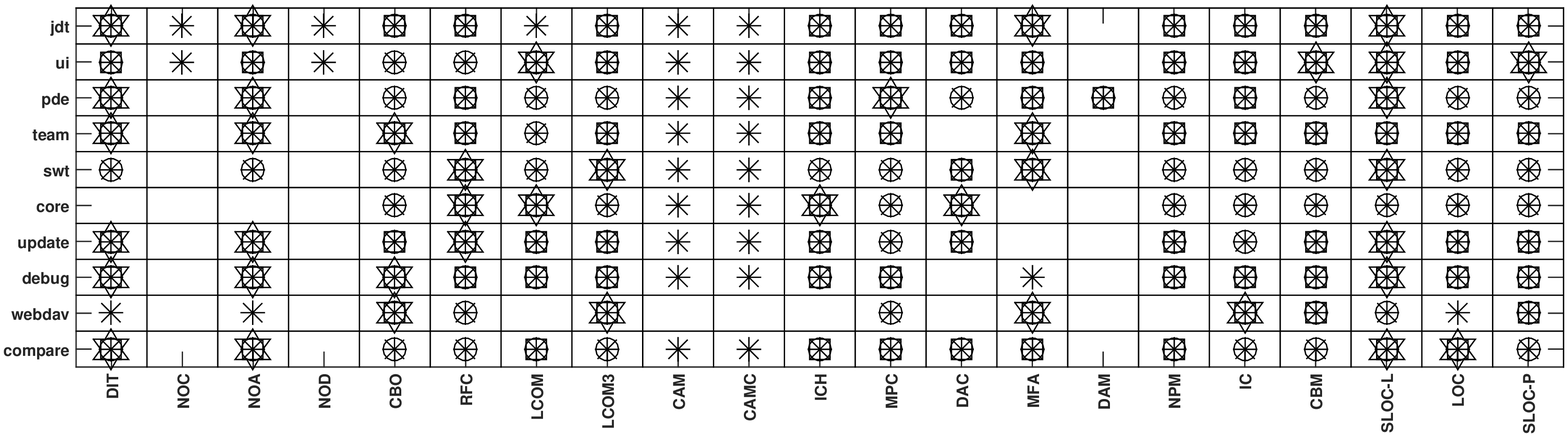}
	\caption{Selected Set of Metrics}
	\label{all4oos}
\end{figure*}

In this work, four different steps have been followed to 
validate the source code metrics. In each steps, some of source code metrics are eventually selected among available source code metrics based on the output of previous steps. Figure \ref{all4oos} shows the selected set of source code metrics in each steps for all considered datasets. The graphs are represented using four different
symbols as described below. The Selected Metrics (SM) are
different for each dataset and displayed using the Hexagonal with square, circle and star in Figure \ref{all4oos}. All Metrics (AM)
comprises of all the 21 metrics shown in Figure \ref{all4oos}.

\begin{itemize}
	\item Star ($\mathlarger{\mathlarger{\ast}}$): source code metrics selected after Wilcoxon signed rank test.
	\item Circle with star ($\mathlarger{\mathlarger{\mathlarger{\mathlarger{\circ}}}}$\hspace{-1.2em} $\mathlarger{\mathlarger{\mathlarger{\ast}}}$): source code metrics selected after Wilcoxon signed rank test and ULR analysis. 
	\item Square with circle and star ($\square$\hspace{-0.8em}$\mathlarger{\mathlarger{\mathlarger{\mathlarger{\circ}}}}$\hspace{-1.1590em} $\mathlarger{\mathlarger{\mathlarger{\ast}}}$): source code metrics selected after Wilcoxon signed rank test, ULR analysis and cross correlation analysis.
	\item Hexagonal with square, circle and star ($\mathlarger{\mathlarger{\varhexagon}}\hspace{-0.74em} \Square$ \hspace{-1.15em}$\mathlarger{\mathlarger{\circ}}$)\hspace{-1.4em} $\mathlarger{\mathlarger{\ast}}$: source code metrics selected after Wilcoxon signed rank test, ULR analysis, cross correlation analysis and MLR stepwise forward selection method.
\end{itemize} 

\subsection{Feature selection methods} 
After finding the right set of source code metrics for change-proneness models using proposed framework, this paper also
makes the comparison of proposed software metrics validation
framework (PFSM) with other ten most frequently used feature selection techniques. Feature selection is a process of selecting a suitable subset of object-oriented metrics from the list of available metrics. Feature selection methods are classified into two subclasses consisting of feature ranking and feature subset selection methods. In feature ranking methods, decisive factors are considered to rank each individual feature and higher ranked features applicable for a given project are chosen. In feature subset selection methods, subset of features are identified to collectively improve predictive capability.

\subsubsection{Feature ranking methods} 
In this study, five different feature ranking methods are applied to 
to eliminate some of the irrelevant and redundant original
variables to increase the training speed and accuracy of the
classifiers. The five different feature ranking methods that are used in this study are (\textbf{(1)} Chi Squared test (FR1), \textbf{(2)} Gain Ratio Feature Evaluation (FR2), \textbf{(3)} oneR Feature Evaluation (FR3), \textbf{(4)} info gain feature evaluation (FR4), \textbf{(5)} principal component analysis (PCA) (FR5). These five methods provides us guidance on selection a subset of the original features which are useful in developing a good estimator or predictor of
change-proneness. Each method uses different performance parameters to rank the features. Further top $\lceil$ $\log_{2}n$$\rceil$ metrics out of 'n' number of metrics have been considered to develop a model for predicting change-proneness. In this study n = 21 and hence top 5 metrics are selected for change-proneness prediction. But in case PCA, only those metrics are selected which have Eigenvalue being more than 1. Table \ref{pcaoos} shows the relation between the original object-oriented metrics and the domain metrics. Values greater then 0.7 (shown bold in Table \ref{pcaoos}) are the object-oriented metrics, which are used to interpret the principal component. Table \ref{pcaoos} also shows the cumulative percentage, eigenvalue, and variance percentage. The metrics selected using ing chi square test are CBO, RFC, MPC, IC, CBM and the 5 metrics selected using gain ratio are RFC, MPC, IC, CBM, LOC for compare dataset.

\begin{table}[h!]
	\caption{Rotated principle component}
	\label{pcaoos}
	\centering
	
	\begin{tabular}{|l|*{5}{c|}r}
		\hline
		
		&\textbf{ PC1 }& \textbf{PC2} & \textbf{PC3} & \textbf{PC4}  \\ \hline
		DIT & 0.14 & \textbf{0.94} & 0.17 & 0.08  \\ \hline
		NOC & 0.08 & 0.03 & 0.12 & \textbf{0.95}  \\ \hline
		NOA & 0.14 & \textbf{0.939} & 0.167 & 0.076  \\ \hline
		NOD & 0.04 & 0.006 & 0.093 & \textbf{0.956}  \\ \hline
		CBO & \textbf{0.8} & 0.202 & 0.295 & 0.07  \\ \hline
		RFC & 0.489 & 0.554 & 0.191 & -0.086  \\ \hline
		LCOM & \textbf{0.962} & 0.029 & 0.004 & 0.082  \\ \hline
		LCOM3 & \textbf{0.734} & 0.053 & 0.353 & 0.141  \\ \hline
		CAM & -0.188 & -0.058 & \textbf{-0.947} & -0.112  \\ \hline
		CAMC & -0.188 & -0.058 & \textbf{-0.947} & -0.113  \\ \hline
		ICH & \textbf{0.956} & 0.005 & 0.009 & -0.029  \\ \hline
		MPC & \textbf{0.961} & 0.159 & 0.043 & -0.017  \\ \hline
		DAC & \textbf{0.835} & 0.11 & 0.26 & 0.073  \\ \hline
		MFA & -0.08 & \textbf{0.75} & -0.162 & -0.102  \\ \hline
		DAM & 0.109 & \textbf{-0.7} & 0.398 & 0.04  \\ \hline
		NPM & \textbf{0.874} & 0.018 & 0.31 & 0.173  \\ \hline
		IC & \textbf{0.968} & 0.107 & 0.029 & -0.019  \\ \hline
		CBM & \textbf{0.963} & 0.151 & 0.039 & -0.017  \\ \hline
		SLOC-L & 0.301 & \textbf{0.82} & 0.212 & 0.109  \\ \hline
		LOC & \textbf{0.98} & 0.047 & 0.085 & 0.032  \\ \hline
		SLOC-P & \textbf{0.98} & 0.048 & 0.085 & 0.021  \\ \hline
		Eigenvalues & 9.66 & 3.925 & 2.532 & 1.956  \\ \hline
		\% variance & 45.99 & 18.69 & 12.05 & 9.31  \\ \hline
		Cumulative \% variance & 45.99 & 64.689 & 76.744 & 86.057  \\ \hline

	\end{tabular}
	
\end{table}	

\subsubsection{Feature subset selection methods}
In this work, five different types of feature subset selection methods are applied over $10$ datasets to reduces dimensionality by selecting a subset of metrics that preserves as much information present in the original set of metrics. The five different feature subset selection methods that are used in this study are \textbf{(1)} correlation based feature selection (CFS) technique (FS1), \textbf{(2)} consistency feature selection (FS2), \textbf{(3)} filtered subset evaluation (FS3), \textbf{(4)} rough set analysis (RSA) (FS4), and \textbf{(5)} genetic algorithm (FS5). We use five different feature subset selection methods. This objective is to determine the suitable set of metrics and then use them as predictors for change proneness. Our analysis and results reveals that the dimensionality of the attributes has been reduced from 21 to 9 for compare project using  correlation based feature selection.

\subsection{Classifiers Performance Evaluation}
In this work, 18 different classification algorithm such as linear regression (LINR), polynomial regression (POLR), logistic regression
(LOGR), decision tree (DT), support vector machine (SVM) with linear kernel (SVM-LIN), SVM with polynomial kernel (SVM-POLY), SVM with RBF kernel (SVM-RBF), extreme learning machine (ELM) with linear kernel (ELM-LIN), ELM with polynomial kernel (ELM-POLY), ELM with RBF kernel (ELM-RBF), least square SVM (LSSVM) with linear kernel (LSSVM-LIN), LSSVM with polynomial kernel (LSSVM-POLY), LSSVM with RBF kernel (LSSVM-RBF), and neural network with five different training algorithms,
normally Gradient descent (GD) method, Gradient descent with momentum (GDM) method, Gradient descent with adaptive learning rate (GDA) method, Quasi-Newton method (NM), and Levenberg Marquardt (LM) method. and 3 different ensemble techniques such as Majority Voting Ensemble (MVE) methods, Nonlinear Ensemble Decision Tree Forest
(NDTF) method, Best Training Ensemble (BTE) and
resulting in 21 different predictive model building approaches have been considered to develop a model for prediction change-proneness of object-oriented software system. The performance of these models are evaluated using accuracy (\%) and F-Measure. In this study,  predicting change-proneness of classes is a binary classification
problem and both accuracy and F-Measure are common evaluation metrics for such problems. We infer the following :
\begin{itemize}
	\item In most of the cases, the model developed by considering selected set of metrics using feature selection
	techniques as input obtained better performance i.e.,
	high vales of accuracy and F-Measure for predicting change-
	proneness as compared to a model developed using all
	metrics.
	\item Least square support vector machine with RBF kernel function (LSSVM-RBF) yields better results when compared to other classification.
	\item Nonlinear Ensemble Decision Tree Forest
	(NDTF) method ensemble method outperformed
	as compared to ensemble methods.
\end{itemize}

In this section, boxplot analysis has been also employed to determine which of the selected set of source code metrics and classification techniques work better for change-proneness prediction. Box-plot diagrams help
to observe performance of all methods based on a single diagram. The line in the middle of each box represents the median value. The model which has high median value is the best model for change-proneness prediction. Table \ref{dstoos} shows the descriptive statistics and box plot of the accuracy and f-measure values for the twenty one classifiers. From Table \ref{dstoos}, it is observed that model developed using LSSVM-RBF have high median value of performance parameters as compare to other classification techniques. From Table \ref{dstoos}, it is also observed that Nonlinear Ensemble Decision Tree Forest
(NDTF) method ensemble method outperformed as compared to all other classifier except LSSVM-RBF.
\begin{table*}[ht]
	\caption{Descriptive Statistics of the Performance of the twenty one Classifiers in-terms of Accuracy and F-Measure }
	\label{dstoos}
	\renewcommand{\arraystretch}{1.1}
	\resizebox{15.5cm}{!}{
		\begin{tabular}{|l|c|c|c|c|c|c|c|c|c|c|c|c|c|c|} \hline	
			& \multicolumn{7}{|c|}{\textbf{Accuracy}} & \multicolumn{7}{|c|}{\textbf{F-Measure}} \\ \hline
			& \textbf{Min} & \textbf{Max} & \textbf{Mean} & \textbf{Median} & \textbf{Std Dev} & \textbf{Q1} & \textbf{Q3}  	& \textbf{Min} & \textbf{Max} & \textbf{Mean} & \textbf{Median} & \textbf{Std Dev} & \textbf{Q1} & \textbf{Q3} \\ \hline
			LINR & 55.38 & 83.65 & 72.01 & 71.08 & 6.52 & 67.67 & 78.16 & 0.518 & 0.884 & 0.721 & 0.693 & 0.101 & 0.634 & 0.835  \\ \hline
			POLYR & 52.44 & 83.65 & 68.39 & 68.07 & 8.77 & 63.20 & 76.00 & 0.517 & 0.883 & 0.717 & 0.679 & 0.103 & 0.622 & 0.830  \\ \hline
			LOGR & 63.77 & 83.65 & 76.29 & 76.75 & 4.23 & 73.56 & 79.68 & 0.104 & 0.889 & 0.722 & 0.702 & 0.122 & 0.660 & 0.835  \\ \hline
			DT & 60.75 & 85.58 & 73.82 & 73.49 & 5.60 & 69.82 & 78.31 & 0.451 & 0.899 & 0.698 & 0.670 & 0.110 & 0.623 & 0.803  \\ \hline
			SVM-LIN & 62.10 & 84.34 & 72.24 & 72.28 & 6.26 & 67.67 & 76.57 & 0.000 & 0.886 & 0.535 & 0.669 & 0.347 & 0.116 & 0.840  \\ \hline
			SVM-POLY & 61.83 & 84.34 & 73.10 & 73.81 & 6.14 & 67.67 & 77.76 & 0.000 & 0.889 & 0.579 & 0.717 & 0.322 & 0.300 & 0.838  \\ \hline
			SVM-RBF & 38.81 & 79.07 & 63.98 & 63.44 & 6.95 & 60.41 & 68.17 & 0.000 & 0.864 & 0.436 & 0.563 & 0.330 & 0.000 & 0.714  \\ \hline
			ELM-LIN & 34.62 & 83.13 & 66.08 & 67.87 & 11.30 & 62.84 & 73.05 & 0.000 & 0.882 & 0.393 & 0.398 & 0.326 & 0.043 & 0.701  \\ \hline
			ELM-POLY & 62.79 & 86.54 & 76.52 & 76.00 & 5.01 & 73.78 & 80.13 & 0.000 & 0.909 & 0.697 & 0.686 & 0.177 & 0.624 & 0.841  \\ \hline
			ELM-RBF & 62.84 & 84.34 & 73.40 & 73.09 & 5.59 & 69.07 & 76.55 & 0.000 & 0.885 & 0.557 & 0.637 & 0.315 & 0.395 & 0.837  \\ \hline
			LSSVM-LIN & 63.36 & 89.42 & 78.55 & 79.16 & 5.52 & 75.03 & 82.73 & 0.027 & 0.928 & 0.693 & 0.733 & 0.206 & 0.604 & 0.863  \\ \hline
			LSSVM-POLY & 65.72 & 100.00 & 81.18 & 81.93 & 6.59 & 75.52 & 85.63 & 0.137 & 1.000 & 0.749 & 0.777 & 0.172 & 0.662 & 0.880  \\ \hline
			LSSVM-RBF & \cellcolor{green!20}72.85 & \cellcolor{green!20}100.00 & \cellcolor{green!20}87.20 & \cellcolor{green!20}86.78 & \cellcolor{green!20}8.37 & \cellcolor{green!20}79.43 & \cellcolor{green!20}95.51 & \cellcolor{green!20}0.599 & \cellcolor{green!20}1.000 & \cellcolor{green!20}0.854 & \cellcolor{green!20}0.885 & \cellcolor{green!20}0.116 & \cellcolor{green!20}0.768 & \cellcolor{green!20}0.943  \\ \hline
			NGD & 51.88 & 82.69 & 69.78 & 69.50 & 5.92 & 65.19 & 74.70 & 0.000 & 0.888 & 0.549 & 0.577 & 0.271 & 0.363 & 0.805  \\ \hline
			NGDM & 50.03 & 85.58 & 67.62 & 68.61 & 7.09 & 62.13 & 72.29 & 0.000 & 0.904 & 0.524 & 0.479 & 0.271 & 0.335 & 0.808  \\ \hline
			NGDA & 52.88 & 83.65 & 68.28 & 67.87 & 6.21 & 63.68 & 72.28 & 0.000 & 0.887 & 0.548 & 0.547 & 0.248 & 0.373 & 0.786  \\ \hline
			NNM & 63.97 & 86.75 & 74.81 & 74.70 & 4.85 & 71.38 & 78.34 & 0.159 & 0.892 & 0.683 & 0.664 & 0.155 & 0.583 & 0.831  \\ \hline
			NLM & 62.65 & 86.54 & 75.16 & 74.79 & 4.73 & 71.88 & 78.60 & 0.329 & 0.910 & 0.698 & 0.678 & 0.133 & 0.629 & 0.822  \\ \hline
			BTE & 60.75 & 100.00 & 80.81 & 78.63 & 11.92 & 70.68 & 95.06 & 0.451 & 1.000 & 0.787 & 0.819 & 0.154 & 0.649 & 0.938  \\ \hline
			MVE & 62.84 & 86.75 & 77.57 & 77.20 & 5.45 & 74.33 & 81.95 & 0.000 & 0.908 & 0.689 & 0.708 & 0.198 & 0.610 & 0.849  \\ \hline
			NDTF & 63.17 & 100.00 & 82.19 & 81.76 & 10.05 & 73.04 & 89.87 & 0.491 & 1.000 & 0.800 & 0.834 & 0.133 & 0.670 & 0.908  \\ \hline

	\end{tabular}}
\end{table*}
\subsection{Feature Selection Methods Performance Evaluation}
In this work, twelve different sets of metrics i.e., AM, 5 sets of metrics using 5 feature ranking methods, 5 sets of metrics using 5 feature subset selection methods, and 1 set of metrics using proposed source code metrics validation framework have been considered for change-proneness prediction. Table \ref{dstfsoos} shows the descriptive statistics and box plot of the accuracy and f-measure values for the different sets of source code metrics. From Figure Table \ref{dstfsoos}, it is observed that the change-proneness prediction model developed using selected set of source code metrics using proposed source code metrics framework (PFST) have high median value of performance parameters as compare to other. From Table \ref{dstfsoos}, it has been also observed that, there exists a small subset of source code software metrics out of total available source code software metrics which are able to predict change-proneness with higher accuracy and reduced value of mis-classified errors.

\begin{table*}[ht]
	\caption{Descriptive Statistics of the Performance of the eleven feature selection technique in-terms of Accuracy and F-Measure }
	\label{dstfsoos}
	\renewcommand{\arraystretch}{1.1}
	\resizebox{15.5cm}{!}{
		\begin{tabular}{|l|c|c|c|c|c|c|c|c|c|c|c|c|c|c|} \hline	
			& \multicolumn{7}{|c|}{\textbf{Accuracy}} & \multicolumn{7}{|c|}{\textbf{F-Measure}} \\ \hline
			& \textbf{Min} & \textbf{Max} & \textbf{Mean} & \textbf{Median} & \textbf{Std Dev} & \textbf{Q1} & \textbf{Q3}  	& \textbf{Min} & \textbf{Max} & \textbf{Mean} & \textbf{Median} & \textbf{Std Dev} & \textbf{Q1} & \textbf{Q3} \\ \hline
			AM & 38.81 & 100.00 & 74.77 & 74.70 & 9.39 & 70.34 & 79.52 & 0.14 & 1.00 & 0.69 & 0.69 & 0.17 & 0.58 & 0.84  \\ \hline
			FR1 & 36.63 & 99.25 & 74.60 & 74.60 & 9.78 & 69.25 & 80.72 & 0.00 & 0.99 & 0.65 & 0.71 & 0.26 & 0.58 & 0.84  \\ \hline
			FR2 & 36.40 & 99.20 & 73.56 & 73.67 & 10.32 & 66.95 & 79.52 & 0.00 & 0.99 & 0.61 & 0.67 & 0.30 & 0.52 & 0.84  \\ \hline
			FR3 & 46.99 & 99.25 & 74.92 & 74.80 & 9.21 & 68.38 & 80.77 & 0.00 & 0.99 & 0.66 & 0.71 & 0.25 & 0.59 & 0.85  \\ \hline
			FR4 & 36.00 & 99.25 & 73.52 & 74.06 & 9.02 & 68.67 & 79.52 & 0.00 & 0.99 & 0.64 & 0.67 & 0.25 & 0.58 & 0.84  \\ \hline
			FR5 & 52.88 & 95.98 & 73.74 & 72.89 & 7.65 & 68.42 & 78.85 & 0.00 & 0.94 & 0.65 & 0.71 & 0.24 & 0.59 & 0.83  \\ \hline
			FS1 & 51.88 & 100.00 & 74.94 & 74.44 & 9.34 & 69.08 & 80.40 & 0.00 & 1.00 & 0.67 & 0.70 & 0.24 & 0.61 & 0.84  \\ \hline
			FS2 & 34.62 & 97.67 & 74.59 & 74.75 & 8.90 & 71.05 & 79.70 & 0.00 & 0.98 & 0.67 & 0.71 & 0.22 & 0.60 & 0.84  \\ \hline
			FS3 & 54.22 & 100.00 & 74.72 & 74.46 & 8.73 & 68.67 & 79.65 & 0.00 & 1.00 & 0.66 & 0.71 & 0.24 & 0.61 & 0.83  \\ \hline
			FS4 & 53.76 & 89.16 & 72.35 & 72.13 & 7.40 & 67.20 & 77.44 & 0.00 & 0.92 & 0.61 & 0.67 & 0.27 & 0.53 & 0.82  \\ \hline
			FS5 & 40.00 & 100.00 & 74.09 & 74.22 & 9.25 & 68.17 & 79.12 & 0.00 & 1.00 & 0.63 & 0.69 & 0.28 & 0.57 & 0.84  \\ \hline
			PFST & \cellcolor{green!20}43.60 & \cellcolor{green!20}98.80 & \cellcolor{green!20}75.05 & \cellcolor{green!20}74.56 & \cellcolor{green!20}9.52 & \cellcolor{green!20}68.64 & \cellcolor{green!20}79.70 & \cellcolor{green!20}0.03 & \cellcolor{green!20}0.99 & \cellcolor{green!20}0.64 & \cellcolor{green!20}0.70 & \cellcolor{green!20}0.26 & \cellcolor{green!20}0.52 & \cellcolor{green!20}0.85  \\ \hline

	\end{tabular}}
\end{table*}

\subsection{Classifier Technique and Feature Selection Method Interaction :}	
\label{sel}
This section focuses on the affect of classification methods over the performance of feature selection techniques. In this study 12 different set of source code metrics and 21 different classification techniques have been implemented for investigation. We infer that for each classification method, different set of source code metrics produce better results.

\subsection{Comparison of results}
In this study, statistical test between different pairs of classifiers and different pairs of set of metrics to investigate which of the classifiers and set of metrics performs best in addition to providing accuracy and f-measure comparison data using Table \ref{dstoos} and Table \ref{dstfsoos}. Table \ref{dstoos}, Table \ref{dstfsoos} shows the descriptive statistics of accuracy and f-measure values but do not present any information on statistical tests or significance tests. In this work, pairwise Wilcoxon test with Bonferroni correction has been considered to determine which of the classifiers and selected sets of source code metrics work better or weather they all perform equally well. 
\subsubsection{Classification Techniques}
\begin{figure*}[h]
	\centering
	\includegraphics[width=15.5cm, height=11cm]{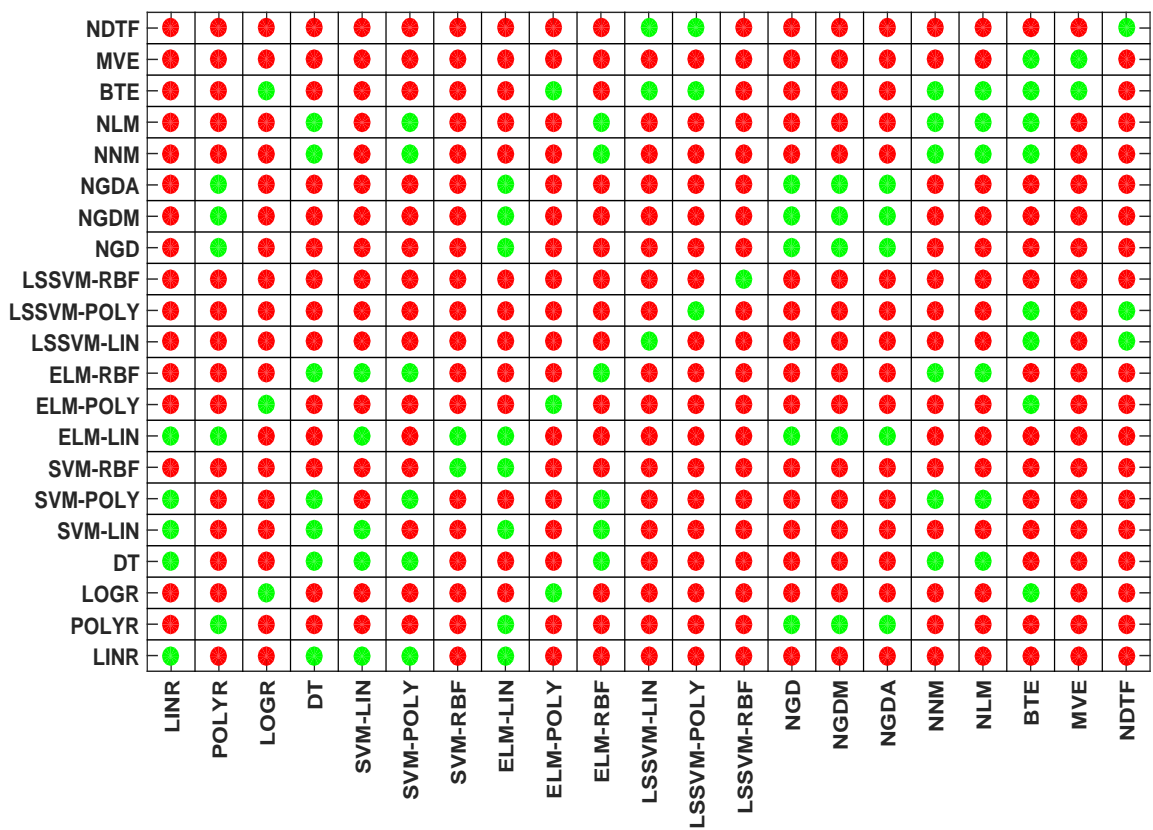}
	\caption{Wilcoxon test with Bonferroni correction (p-value)}
	\label{ttstoos}
\end{figure*}

Eighteen different classification techniques and three different ensemble methods have been considered to develop a model to predict change-proneness classes of object-oriented software. Twelve set of metrics are considered as input to develop a model to predict change-proneness of object-oriented software. Hence, for each classifier, total number of two set (one for each
performance measure) are used, each with 120 data points (12 set of metrics multiplied by 10 datasets). 
In this work, the null hypothesis is that there is no effect on the change-proneness prediction accuracy due to the classifiers i.e., the model developed by all classifiers are similar. In this experiments, initially the p-value is set to $0.05$ then the Bonferroni correction adjust the significance cutoff at $\frac{0.05}{n}$, where $n$ is number of different pairs (in our case it is $21$ classifiers: n=$^{21 technique}C_{2}=21*20/2=210$). Hence, the null hypothesis is accepted only when p-value is less then $\frac{0.05}{210}=0.0002381$. Figure \ref{ttstoos} shows the the pair-wise comparisons of different classifiers. The graphs are represented using two different
symbols such as \textcolor{green}{$\CIRCLE$}: (no significance difference) and \textcolor{red}{$\CIRCLE$}: (significance difference). 
Figure \ref{ttstoos} reveals that there is a significant difference between classifiers in most of the cases due to the fact that the p-value is smaller than 0.0002381, i.e. out of 210 pairs of classifiers, 172 are found to have significant results. From Table \ref{tstoos}, it is observed that the LSSVM-RBF have outperformed when compared to other classifiers based on mean difference of performance parameter.

\begin{table*}[h]
	\caption{Mean Difference between Performance of Different classifiers}
	\label{tstoos}
	\centering
	
	\renewcommand{\arraystretch}{1.1}
	\resizebox{15.5cm}{!}{
		\begin{tabular}{|l|*{22}{c|}r}
			\hline
			\multicolumn{22}{|c|}{\textbf{Accuracy}} \\ \hline
			
			& \rotatebox{80}{\textbf{LINR}} & \rotatebox{80}{\textbf{POLYR}} & \rotatebox{80}{\textbf{LOGR}} & \textbf{DT} & \rotatebox{80}{\textbf{SVM-LIN}} & \rotatebox{80}{\textbf{SVM-POLY}} & \rotatebox{80}{\textbf{SVM-RBF}} & \rotatebox{80}{\textbf{ELM-LIN}} & \rotatebox{80}{\textbf{ELM-POLY}} & \rotatebox{80}{\textbf{ELM-RBF}} & \rotatebox{80}{\textbf{LSSVM-LIN}} & \rotatebox{80}{\textbf{LSSVM-POLY}} & \rotatebox{80}{\textbf{LSSVM-RBF}} & \textbf{NGD} & \textbf{NGDM} & \textbf{NGDA} & \textbf{NNM} & \textbf{NLM} & \textbf{BTE} & \textbf{MVE} & \textbf{NDTF}  \\ \hline
			LINR & 0.00 & 3.62 & -4.27 & -1.80 & -0.22 & -1.09 & 8.03 & 5.93 & -4.50 & -1.38 & -6.54 & -9.17 & -15.18 & 2.23 & 4.39 & 3.74 & -2.79 & -3.14 & -8.80 & -5.56 & -10.18 \\ \hline
			POLYR & -3.62 & 0.00 & -7.89 & -5.42 & -3.84 & -4.70 & 4.41 & 2.32 & -8.12 & -5.00 & -10.16 & -12.79 & -18.80 & -1.39 & 0.77 & 0.12 & -6.41 & -6.76 & -12.42 & -9.18 & -13.80 \\ \hline
			LOGR & 4.27 & 7.89 & 0.00 & 2.47 & 4.05 & 3.19 & 12.30 & 10.21 & -0.23 & 2.89 & -2.27 & -4.90 & -10.91 & 6.50 & 8.67 & 8.01 & 1.48 & 1.13 & -4.52 & -1.28 & -5.90 \\ \hline
			DT & 1.80 & 5.42 & -2.47 & 0.00 & 1.58 & 0.72 & 9.83 & 7.74 & -2.70 & 0.42 & -4.74 & -7.37 & -13.38 & 4.03 & 6.20 & 5.54 & -0.99 & -1.34 & -7.00 & -3.76 & -8.37 \\ \hline
			SVM-LIN & 0.22 & 3.84 & -4.05 & -1.58 & 0.00 & -0.86 & 8.25 & 6.16 & -4.28 & -1.16 & -6.32 & -8.95 & -14.96 & 2.45 & 4.61 & 3.96 & -2.57 & -2.92 & -8.58 & -5.34 & -9.96 \\ \hline
			SVM-POLY & 1.09 & 4.70 & -3.19 & -0.72 & 0.86 & 0.00 & 9.11 & 7.02 & -3.42 & -0.30 & -5.46 & -8.09 & -14.10 & 3.31 & 5.48 & 4.82 & -1.71 & -2.06 & -7.71 & -4.47 & -9.09 \\ \hline
			SVM-RBF & -8.03 & -4.41 & -12.30 & -9.83 & -8.25 & -9.11 & 0.00 & -2.09 & -12.53 & -9.41 & -14.57 & -17.20 & -23.21 & -5.80 & -3.64 & -4.29 & -10.82 & -11.17 & -16.83 & -13.59 & -18.21 \\ \hline
			ELM-LIN & -5.93 & -2.32 & -10.21 & -7.74 & -6.16 & -7.02 & 2.09 & 0.00 & -10.44 & -7.32 & -12.48 & -15.11 & -21.12 & -3.71 & -1.54 & -2.20 & -8.73 & -9.08 & -14.73 & -11.49 & -16.11 \\ \hline
			ELM-POLY & 4.50 & 8.12 & 0.23 & 2.70 & 4.28 & 3.42 & 12.53 & 10.44 & 0.00 & 3.12 & -2.04 & -4.67 & -10.68 & 6.73 & 8.89 & 8.24 & 1.71 & 1.36 & -4.30 & -1.06 & -5.68 \\ \hline
			ELM-RBF & 1.38 & 5.00 & -2.89 & -0.42 & 1.16 & 0.30 & 9.41 & 7.32 & -3.12 & 0.00 & -5.16 & -7.79 & -13.80 & 3.61 & 5.77 & 5.12 & -1.41 & -1.76 & -7.42 & -4.18 & -8.80 \\ \hline
			LSSVM-LIN & 6.54 & 10.16 & 2.27 & 4.74 & 6.32 & 5.46 & 14.57 & 12.48 & 2.04 & 5.16 & 0.00 & -2.63 & -8.64 & 8.77 & 10.93 & 10.28 & 3.75 & 3.40 & -2.26 & 0.98 & -3.64 \\ \hline
			LSSVM-POLY & 9.17 & 12.79 & 4.90 & 7.37 & 8.95 & 8.09 & 17.20 & 15.11 & 4.67 & 7.79 & 2.63 & 0.00 & -6.01 & 11.40 & 13.56 & 12.91 & 6.38 & 6.03 & 0.37 & 3.61 & -1.01 \\ \hline
			LSSVM-RBF & \cellcolor{green!20}15.18 & \cellcolor{green!20}18.80 & \cellcolor{green!20}10.91 & \cellcolor{green!20}13.38 & \cellcolor{green!20}14.96 & \cellcolor{green!20}14.10 & \cellcolor{green!20}23.21 & \cellcolor{green!20}21.12 & \cellcolor{green!20}10.68 & \cellcolor{green!20}13.80 & \cellcolor{green!20}8.64 & \cellcolor{green!20}6.01 & \cellcolor{green!20}0.00 & \cellcolor{green!20}17.41 & \cellcolor{green!20}19.57 & \cellcolor{green!20}18.92 & \cellcolor{green!20}12.39 & \cellcolor{green!20}12.04 & \cellcolor{green!20}6.38 & \cellcolor{green!20}9.62 & \cellcolor{green!20}5.00 \\ \hline
			NGD & -2.23 & 1.39 & -6.50 & -4.03 & -2.45 & -3.31 & 5.80 & 3.71 & -6.73 & -3.61 & -8.77 & -11.40 & -17.41 & 0.00 & 2.16 & 1.51 & -5.02 & -5.37 & -11.03 & -7.79 & -12.41 \\ \hline
			NGDM & -4.39 & -0.77 & -8.67 & -6.20 & -4.61 & -5.48 & 3.64 & 1.54 & -8.89 & -5.77 & -10.93 & -13.56 & -19.57 & -2.16 & 0.00 & -0.65 & -7.19 & -7.53 & -13.19 & -9.95 & -14.57 \\ \hline
			NGDA & -3.74 & -0.12 & -8.01 & -5.54 & -3.96 & -4.82 & 4.29 & 2.20 & -8.24 & -5.12 & -10.28 & -12.91 & -18.92 & -1.51 & 0.65 & 0.00 & -6.53 & -6.88 & -12.54 & -9.30 & -13.91 \\ \hline
			NNM & 2.79 & 6.41 & -1.48 & 0.99 & 2.57 & 1.71 & 10.82 & 8.73 & -1.71 & 1.41 & -3.75 & -6.38 & -12.39 & 5.02 & 7.19 & 6.53 & 0.00 & -0.35 & -6.01 & -2.77 & -7.38 \\ \hline
			NLM & 3.14 & 6.76 & -1.13 & 1.34 & 2.92 & 2.06 & 11.17 & 9.08 & -1.36 & 1.76 & -3.40 & -6.03 & -12.04 & 5.37 & 7.53 & 6.88 & 0.35 & 0.00 & -5.66 & -2.42 & -7.04 \\ \hline
			BTE & 8.80 & 12.42 & 4.52 & 7.00 & 8.58 & 7.71 & 16.83 & 14.73 & 4.30 & 7.42 & 2.26 & -0.37 & -6.38 & 11.03 & 13.19 & 12.54 & 6.01 & 5.66 & 0.00 & 3.24 & -1.38 \\ \hline
			MVE & 5.56 & 9.18 & 1.28 & 3.76 & 5.34 & 4.47 & 13.59 & 11.49 & 1.06 & 4.18 & -0.98 & -3.61 & -9.62 & 7.79 & 9.95 & 9.30 & 2.77 & 2.42 & -3.24 & 0.00 & -4.62 \\ \hline
			NDTF & 10.18 & 13.80 & 5.90 & 8.37 & 9.96 & 9.09 & 18.21 & 16.11 & 5.68 & 8.80 & 3.64 & 1.01 & -5.00 & 12.41 & 14.57 & 13.91 & 7.38 & 7.04 & 1.38 & 4.62 & 0.00 \\ \hline

	\end{tabular}}
	
\end{table*}	

\subsubsection{Feature Selection Techniques:}
\begin{figure*}[h]
	\centering
	\includegraphics[width=12.5cm, height=7cm]{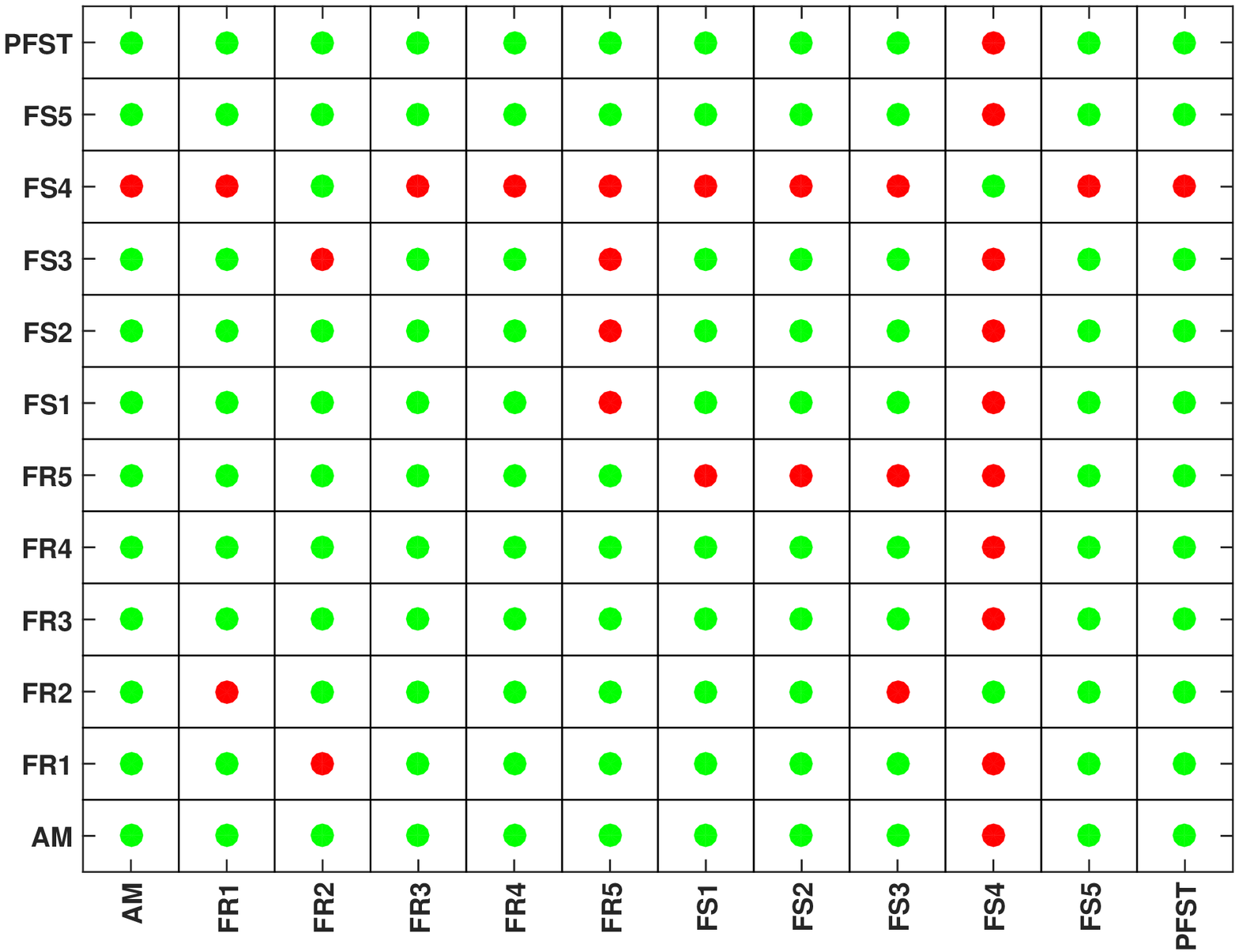}
	\caption{Wilcoxon test with Bonferroni correction (p-value)}
	\label{ttstfsoos}
\end{figure*}

Twelve set of metrics have been considered as input to develop a model to predict change-proneness classes of object-oriented software. Twenty one classifiers are considered to develop a model to predict change-proneness of object-oriented software.  Hence, for set of metrics, total number of two set (one for each
performance measure) are used, each with 210 data points (21 classifiers multiplied by 10 datasets). Figure \ref{ttstfsoos} shows the the pair-wise comparisons of different classifiers. The graphs are represented using two different symbols such as \textcolor{green}{$\CIRCLE$}: (no significance difference) and \textcolor{red}{$\CIRCLE$}: (significance difference). In this work, all results analyzed at 0.05 significance level. Hence, we can only accept a null hypothesis if the p-value is greater $\frac{0.05}{66}=0.0007575$ (Total number of different pairs= $^{12 technique}C_{2}=12*11/2=66$). Figure \ref{ttstfsoos} reveals that there is a no any significant difference between different set of metrics in most of the cases due to the fact that the p-value is greater than 0.0007575, i.e. out of 66 pairs of classifiers, 51 are found to have no significant results. From Table \ref{tstfsoos}, it is observed that the the selected set of metrics using proposed framework (PFST) outperformed when compared to other feature selection technique and also all metrics based on mean difference of performance parameter.

\begin{table*}[h]
	\caption{Mean Difference between Performance of Different feature selection techniques}
	\label{tstfsoos}
	\centering
	
	\renewcommand{\arraystretch}{1.1}
	\resizebox{15.5cm}{!}{
		\begin{tabular}{|l|*{13}{c|}r}
			\hline
			\multicolumn{13}{|c|}{\textbf{Accuracy}} \\ \hline
			
			& \textbf{AM} & \textbf{FR1} & \textbf{FR2} & \textbf{FR3} & \textbf{FR4} & \textbf{FR5} & \textbf{FS1} & \textbf{FS2} & \textbf{FS3} & \textbf{FS4} & \textbf{FS5} & \textbf{PFST}  \\ \hline
			AM  & 0.00  & 0.17  & 1.22  & -0.15  & 1.25  & 1.03  & -0.17  & 0.19  & 0.05  & 2.42  & 0.68  & -0.28  \\ \hline
			FR1  & -0.17  & 0.00  & 1.05  & -0.32  & 1.08  & 0.86  & -0.34  & 0.02  & -0.11  & 2.25  & 0.52  & -0.44  \\ \hline
			FR2  & -1.22  & -1.05  & 0.00  & -1.37  & 0.04  & -0.19  & -1.38  & -1.03  & -1.16  & 1.20  & -0.53  & -1.49  \\ \hline
			FR3  & 0.15  & 0.32  & 1.37  & 0.00  & 1.40  & 1.18  & -0.02  & 0.34  & 0.20  & 2.57  & 0.83  & -0.13  \\ \hline
			FR4  & -1.25  & -1.08  & -0.04  & -1.40  & 0.00  & -0.22  & -1.42  & -1.07  & -1.20  & 1.17  & -0.57  & -1.53  \\ \hline
			FR5  & -1.03  & -0.86  & 0.19  & -1.18  & 0.22  & 0.00  & -1.19  & -0.84  & -0.97  & 1.39  & -0.34  & -1.30  \\ \hline
			FS1  & 0.17  & 0.34  & 1.38  & 0.02  & 1.42  & 1.19  & 0.00  & 0.35  & 0.22  & 2.59  & 0.85  & -0.11  \\ \hline
			FS2  & -0.19  & -0.02  & 1.03  & -0.34  & 1.07  & 0.84  & -0.35  & 0.00  & -0.13  & 2.23  & 0.50  & -0.46  \\ \hline
			FS3  & -0.05  & 0.11  & 1.16  & -0.20  & 1.20  & 0.97  & -0.22  & 0.13  & 0.00  & 2.37  & 0.63  & -0.33  \\ \hline
			FS4  & -2.42  & -2.25  & -1.20  & -2.57  & -1.17  & -1.39  & -2.59  & -2.23  & -2.37  & 0.00  & -1.74  & -2.70  \\ \hline
			FS5  & -0.68  & -0.52  & 0.53  & -0.83  & 0.57  & 0.34  & -0.85  & -0.50  & -0.63  & 1.74  & 0.00  & -0.96  \\ \hline
			PFST  & \cellcolor{green!20}0.28  & \cellcolor{green!20}0.44  & \cellcolor{green!20}1.49  & \cellcolor{green!20}0.13  & \cellcolor{green!20}1.53  & \cellcolor{green!20}1.30  & \cellcolor{green!20}0.11  & \cellcolor{green!20}0.46  & \cellcolor{green!20}0.33  & \cellcolor{green!20}2.70  & \cellcolor{green!20}0.96  & \cellcolor{green!20}0.00  \\ \hline

	\end{tabular}}
	
\end{table*}

\section{Answers to Research Questions}
Based on this study, this work answers the following
research questions.
\begin{itemize}
	\item [RQ1] In this work, Wilcoxon signed rank test and ULR analysis on each source code metrics to investigate the significant association with class change-proneness. According to all datasets results depicted using green circle (\textcolor{green}{$\CIRCLE$}) and red circle (\textcolor{red}{$\CIRCLE$}) in Figure \ref{hypothesisoos}. From Figure \ref{hypothesisoos}, it can be 
	infer that some source code metrics significantly differentiate the change or non-change-proneness classes. From this analysis, it can be observed that source code metrics were significantly correlated with change-proneness.
	\item [RQ2] Table \ref{dstfsoos} reveals  that there exists a reduced subset of object-oriented metrics that is better for designing a prediction model as compared to considering all metrics i.e., model developed by considering selected set of source code metrics using proposed framework outperform as compare to the all metrics. 
	\item [RQ3] The change-proneness model developed using different classifiers are significantly different based on Wilcoxon signed rank test analysis. But, based on the values of mean difference of performance parameters, LSSVM-RBF  yields better results compared to other classifiers.
	
	\item [RQ4] The change-proneness model developed by considering selected set of metrics using feature selection techniques are not significantly different based on Wilcoxon signed rank test analysis. But upon judging the value of mean difference of performance parameters, selected set of source code metrics using proposed framework outperform as compare to the other feature selection techniques. 
	
	\item [RQ5]  Table \ref{dstoos} reveals  that the Nonlinear Ensemble Decision Tree Forest
	(NDTF) method ensemble method outperformed as compared to all other classifier except LSSVM-RBF.
	\item [RQ6] We conclude that the performance of the classifiers varies with the different set of source code metrics. This result shows that selection of classification metrics to develop a model for predicting change-proneness classes is affected by the selection of source code metrics.
\end{itemize}

\section{Summary}
\label{conculsionoos}
This work proposes a comparative study of different set
source code metrics for change-proneness prediction. The
effectiveness of these set of source code metrics are evaluated using eighteen different classifiers and three different ensemble methods. The objective of this study is to investigate the ability of source code metrics to predict change-prone classes. The main observations are the following:
\begin{itemize}
	\item It is possible to accurately predict the change-proneness of object-oriented software using source code metrics.
	\item From experimental results, it is observed that, there
	exists a small subset of source code software metrics
	out of total available source code software metrics which
	are able to predict change-proneness with higher accuracy and reduced value of misclassified errors.
	\item It is observed that the model developed
	using proposed feature selection framework (PFST) yields better result compared to other approaches.
	\item From experimental results, it is observed that model developed using LSSVM-RBF yields better result as compare to
	other classification techniques. From this results, it is also
	observed that Nonlinear Ensemble Decision Tree Forest
	(NDTF) method ensemble method outperformed
	outperformed as compared to all other classifiers except
	LSSVM-RBF.
	\item From experiments, it is observed that the performance
	of the feature selection techniques is varied with the
	different classification methods used.
\end{itemize}

\bibliographystyle{acm}
\bibliography{sigproc}
\end{document}